# Strong coupling regime in two-dimensional large-$N$ scalar quantum chromodynamics

Vladimir Visnic
*Department of Physics, Temple University*
*Philadelphia, PA 19122*

## Abstract

I consider two-dimensional large-$N$ quantum chromodynamics with scalar quarks with particular emphasis on its strong coupling regime which has not been studied so far. Techniques necessary to deal with the infinitely oscillatory bound state wave functions in the strong coupling regime are developed. I derive an estimate for the ground state mass and show that (1) the lightest hadron in the theory is massless and (2) the ground state mass is continuous across the transition between the weak and the strong coupling.

PACS numbers: 02.30.Rz, 11.10.Ef, 11.10.Kk, 11.10.St, 11.15.Pg

Two-dimensional quantum chromodynamics with scalar quarks is defined by the Lagrangian density

$$\mathcal{L} = -\frac{1}{4}F^j_{\mu\nu i}F^{\mu\nu i}_j + (D_\mu\phi)^*_i(D_\mu\phi)^i - m^2\phi^*_i\phi^i. \tag{1}$$

Here the scalar field $\phi_i$ belongs to the fundamental representation of the SU($N$) gauge group, $g$ and $m$ are positive constants having dimension of mass and the covariant derivative is $D_\mu\phi_i = \partial\phi_i + igA^j_{\mu i}\phi_j$. For simplicity and symmetry I consider the theory with one quark flavor. The large-$N$ theory is obtained by taking the limit $N \to \infty$ and $g \to 0$ while keeping $g^2 N = $ constant.

The theory with fermionic quarks was developed by 't Hooft [1] who obtained an integral equation for the quark–antiquark bound state. The corresponding equation in the scalar theory can be obtained by basically retracing his derivation. In the light cone gauge $A_- = 0$ the quark-gluon vertex is $g(p_1 + p_2)$ while the quartic coupling $g^2 A^2 \phi^* \phi$ vanishes. In the large-$N$ limit the dressed quark propagator is

$$\frac{i}{2k_+k_- - m^2 - g^2N/\pi - g^2|k_-|/\pi\lambda + i\epsilon},$$

where $\lambda$ is the infrared cutoff. This leads to the following equation for the bound state of the scalar quark–antiquark pair:

$$M^2\phi(x) = \left(\frac{m^2 - \frac{g^2N}{\pi}}{x(1-x)} + \frac{g^2N}{4\sqrt{x(1-x)}}\right)\phi(x)$$
$$+ \frac{g^2N}{\pi}\int_0^1 \frac{\phi(x) - \phi(y)}{(x-y)^2} \frac{(x+y)(2-x-y)}{4\sqrt{x(1-x)y(1-y)}}dy, \tag{2}$$

where $M$ is the mass of the quark–antiquark bound state, $\phi$ is the bound state wave function, and $x$ is the light cone momentum scaling variable. This equation was first derived by Bardeen *et al.* [2] and subsequently also studied by other authors [3,4].

In the above references attention was restricted to the weak coupling or, equivalently, heavy quark regime of the large-$N$ theory. In this Letter particular attention will be paid to the strong coupling regime of the theory, although some weak coupling results will be derived as well. It turns out that the theory exhibits markedly different behavior in its strong and weak coupling regimes. In particular, the bound state wave functions in the two regimes have entirely different properties. I will derive an approximate analytic expression for the ground state mass as a function of the quark mass and show that, contrary to what was previously believed, the lightest meson in the theory is massless. I also show that the mass of the ground state changes continuously as we pass from one regime to the other thus indicating that the spectrum of the theory depends continuously on the coupling (or quark mass).

First note that there is an equivalence between light quarks and strong coupling (and between heavy quarks and weak coupling) in the large-$N$ theory. The equivalence is based on the fact that the bound state equation (2) involves a dimensionless parameter, $\frac{g^2 N}{\pi m^2}$. When $\frac{g^2 N}{\pi m^2} < 1$ we have the weak coupling/heavy quark regime and when $\frac{g^2 N}{\pi m^2} > 1$ we have the strong coupling/light quark regime. In the limit of infinite quark mass the theory describes free scalar particles and can be exactly solved. The bound state equation reduces to

$$M^2 \phi(x) = m^2 \frac{\phi(x)}{x(1-x)}$$

whose solutions are $\phi(x) = \delta(x - \frac{1}{2})$ and $M = 2m$. In the limit $\frac{g^2 N}{\pi m^2} \to \infty$, we have massless or, equivalently, infinitely strongly coupled QCD. As I will show the theory can be almost exactly solved in this limit and the ground state is massless.

The first step in tackling the bound state equation (2) is to identify the proper Hilbert spaces in both regimes. This is done by studying the boundary conditions that the wave function must satisfy such that the leading singularities on the right hand side of Eq. (2) cancel. In the weak coupling regime the coefficient of the kinetic term is positive and the singularity cancellation dictates that the wave function must vanish at the boundary points 0 and 1 as $x^\beta$ and $(1-x)^\beta$; here $0 < \beta < 1/2$ depends on the mass and the coupling constant as follows:

$$m^2 - \frac{g^2 N}{\pi} = \frac{g^2 N}{\pi} \pi \beta \tan \pi \beta. \tag{3}$$

This equation maps the entire weak coupling regime, $0 < \frac{g^2 N}{\pi m^2} < 1$, onto the entire available range of $\beta$ between 0 and 1/2 and has no solution in the strong coupling regime. That is, it has no *real* solution. I will adopt here the rather unorthodox point of view of allowing imaginary exponents. Indeed, one can show that the imaginary exponent wave functions $x^{i\beta}$ and $(1-x)^{i\beta}$ lead to the required singularity cancellations when $\frac{g^2N}{\pi m^2} > 1$. In this case the equation for $\beta$ is

$$\frac{g^2 N}{\pi} - m^2 = \frac{g^2 N}{\pi}\pi\beta\tanh\pi\beta. \tag{4}$$

This equation maps the entire strong coupling regime $1 < \frac{g^2N}{\pi m^2} < \infty$ onto the imaginary-$\beta$ axis between 0 and 0.38187... Hence the wave functions of hadrons made up of massless quarks must behave at the boundaries as $x^{i\,0.38187...}$ and $(1-x)^{i\,0.38187...}$.

I conclude that in the strong coupling regime the Hilbert space consists of $L^2(0,1)$ functions of the form

$$\phi(x) = A(x)x^{i\beta}(1-x)^{i\beta}, \tag{5}$$

with $A(x)$ behaving near the boundary as $x^0$ and $(1-x)^0$. Taking into account the symmetry of the equation, $A(x)$ can be shown to be either of the form

$$A(x) = \sum_{k=0}^{\infty} a_k x^k (1-x)^k, \quad a_0 \neq 0, \tag{6}$$

or of the form

$$A(x) = (1-2x)\sum_{k=0}^{\infty} a_k x^k (1-x)^k, \quad a_0 \neq 0^\star. \tag{7}$$

The physics of the imaginary exponent wave functions is worth commenting on. When the quarks are heavy and the coupling is weak, the probability of finding them near the boundary, i.e. at very large relative momenta, is vanishingly small. They deal with the singularities by staying away from the regions where the singularities are. Light and tightly bound quarks, on the contrary, have finite probability to be found near the boundary. The way they kill the singularities is by having wave functions which oscillate infinitely rapidly in the singular regions!

The Hamiltonian matrix element between the states $\psi$ and $\phi$ is

$$(\psi, H\phi) = (m^2 - \frac{g^2 N}{\pi})\int_0^1 \frac{\psi(x)^*\phi(x)}{x(1-x)}dx + \frac{1}{4}g^2 N \int_0^1 \frac{\psi(x)^*\phi(x)}{\sqrt{x(1-x)}}dx$$

$$+\frac{g^2 N}{\pi}\int_0^1\int_0^1 \frac{(\psi^*(x)-\psi^*(y))(\phi(x)-\phi(y))}{(x-y)^2}\frac{(x+y)(2-x-y)}{8\sqrt{x(1-x)y(1-y)}}dxdy. \quad (8)$$

The functions (5) have unusual properties and the evaluation of the integrals is subtle. For example, the absolute value of $A(x)x^{i\beta}(1-x)^{i\beta}$ is independent of $\beta$ and equal to $A(x)^2$, even though both the real and the imaginary part, $A(x)\cos(\beta \ln x(1-x))$ and $A(x)\sin(\beta \ln x(1-x))$, are highly oscillatory near the boundaries. In general, in the product $\psi^*(x)\phi(x)$ there is no vestige of the highly oscillatory behavior of its factors: it goes smoothly to non-zero values at the boundary points. This, of course, leaves the singularities in the kinetic term out of control. It is less obvious, but equally true, that double integrals involving products of the wave functions at different points $\psi^*(x)\phi(y)$ also blow up for the same reason. Apparently, even though the boundary behavior of $\phi(x)$ cancels the singularities in the integral equation (2), it does not do the same in the Hamiltonian matrix element $(\psi, H\phi)$. This is unphysical as we cannot allow the Hamiltonian matrix elements to have singularities which are not present in the eigenvalue equation (2).

To remove the singularities I propose an alternative definition of the product $\psi^*(x)\phi(y)$ which will "remember" that $\phi$ and $\psi$ are highly oscillatory functions. The following prescription gives correct and physically meaningful results: replace the product $\psi^*(x)\phi(y)$ by

$$(\psi^*(x)\phi(y))_{reg} = \frac{1}{2}(\psi^*(x)\tilde{\phi}(y) + \tilde{\psi}^*(x)\phi(y)), \quad (9)$$

where

$$\begin{aligned}\tilde{\phi}(x) &= \frac{1}{2}\phi(x)(x^{i\epsilon}(1-x)^{i\epsilon} + x^{-i\epsilon}(1-x)^{-i\epsilon}) \\ &= \frac{1}{2}A(x)(x^{i(\beta+\epsilon)}(1-x)^{i(\beta+\epsilon)} + x^{i(\beta-\epsilon)}(1-x)^{i(\beta-\epsilon)})\end{aligned} \quad (10)$$

and let $\epsilon \to 0$ at the end of the calculation. Note that since the tilde regularization commutes with the complex conjugation, the Hamiltonian remains hermitian under the regularization. With this prescription the integrals are well defined and can be computed.

In both cases (6) and (7) it is the constant term which leads to the singular integrands in Eq. (8). In other words, if we can establish that the matrix element $(\phi, H\phi)$ exists for the first order wave function $\phi(x) = x^{i\beta}(1-x)^{i\beta}$, the higher order terms will not pose any new problems. The first order wave function is also relevant for the question of positive definiteness of the Hamiltonian and will provide us with an estimate of the ground state mass. Here I present the first order calculation. Higher order corrections and calculation of excited states will be reported elsewhere.

One may wonder whether the Hamiltonian is positive definite in the strong coupling regime since the kinetic term is proportional to the quantity $m^2 - \frac{g^2 N}{\pi}$ which is negative. However, the kinetic term integral turns out to be zero for the wave function $\phi(x) = x^{i\beta}(1-x)^{i\beta}$ for all $\beta$, i.e. throughout the strong coupling regime! This is a direct consequence of the regularization method (9) and it eminently makes sense. Note that as we approach the strong–weak boundary from the weak side, the kinetic term decreases and reaches the value zero at the boundary because its coefficient, the renormalized mass squared, becomes zero. From then on the kinetic term stays zero throughout the strong coupling regime – but now because the integral is zero. Since the remaining terms in (8) are manifestly positive the Hamiltonian is positive definite.

The double integrals are evaluated by first regularizing them with the tilde products (9), computing the integrals, expanding the result in $\epsilon$ and then letting $\epsilon \to 0$. The result can be expressed in closed form in terms of Euler's psi function:

$$
\begin{aligned}
(\phi, H\phi) &= \frac{g^2 N}{\pi} \pi \tanh \pi\beta \\
&\times \left[ \frac{1}{2}\beta + \operatorname{Im}\left\{ (\frac{1}{2} - i\beta)\Big(\psi(-1+2i\beta) + \psi(1-2i\beta) - \psi(-\frac{1}{2}+i\beta) - \psi(\frac{1}{2}-i\beta)\Big) \right\} \right]
\end{aligned}
$$

(12)

This equation is also an approximate formula (more precisely, an upper bound) for the mass squared of the ground state meson. Together with $m^2(\beta)$ given in Eq. (4) it defines parametrically the mass of the ground state as a function of the quark mass. This function is shown in Fig. 1, to the left of the point $m^2 = \frac{g^2 N}{\pi}$.

When the quarks are massless, the upper bound (11) for the ground state mass squared is 0.08 (in units $\frac{g^2 N}{\pi}$), leaving little doubt that the ground state is exactly massless and its wave function very nearly

$$\phi(x) = (x(1-x))^{i\,0.38187...}. \tag{13}$$

It is amusing to observe that the absolute value of this wave function is equal to the exact ground state wave function of massless quarks in the fermion theory, namely 1.

In order to compare these results with the weak coupling theory, I derive now the weak coupling regime analogue of Eq. (11). The Hilbert space now consists of the functions of the form

$$A(x)x^\beta(1-x)^\beta, \tag{14}$$

with $A(x)$ behaving near the boundary points as $x^0$ and $(1-x)^0$. Let us do again the first order calculation of the Hamiltonian matrix element which will give us the ground state mass as a function of $\beta$. The double integrals can be expressed in terms of $\Gamma$ functions by using an integral formula given in Ref. [2] and another formula derived by employing the Wilf–Zeilberger recurrence-finding algorithm [5]. The resulting Hamiltonian matrix element is

$$(\phi, H\phi) = \frac{g^2 N}{\pi} \frac{\Gamma(2+4\beta)}{\Gamma(1+2\beta)^2}$$
$$\left( \pi\beta \tan \pi\beta \frac{\Gamma(2\beta)^2}{\Gamma(4\beta)} + \frac{2\beta+1}{4} \frac{\Gamma(\frac{1}{2}+\beta)^4}{\Gamma(1+2\beta)^2} + \pi \frac{\Gamma(\frac{1}{2}+\beta)^2}{2^{1+4\beta}\Gamma(\beta)\Gamma(1+\beta)} \right) \tag{15}$$

Together with $m^2(\beta)$ in Eq. (3) this expression defines parametrically the mass of the ground state as a function of the quark mass in the weak coupling regime. In Fig. 1, this curve is to the right of the point $m^2 = \frac{g^2 N}{\pi}$. The fact that both formulas join continuously at the point $m^2 = \frac{g^2 N}{\pi}$ where the strong and weak coupling regimes meet is a nice confirmation that our procedure for dealing with divergent integrals in the strong coupling regime is correct.

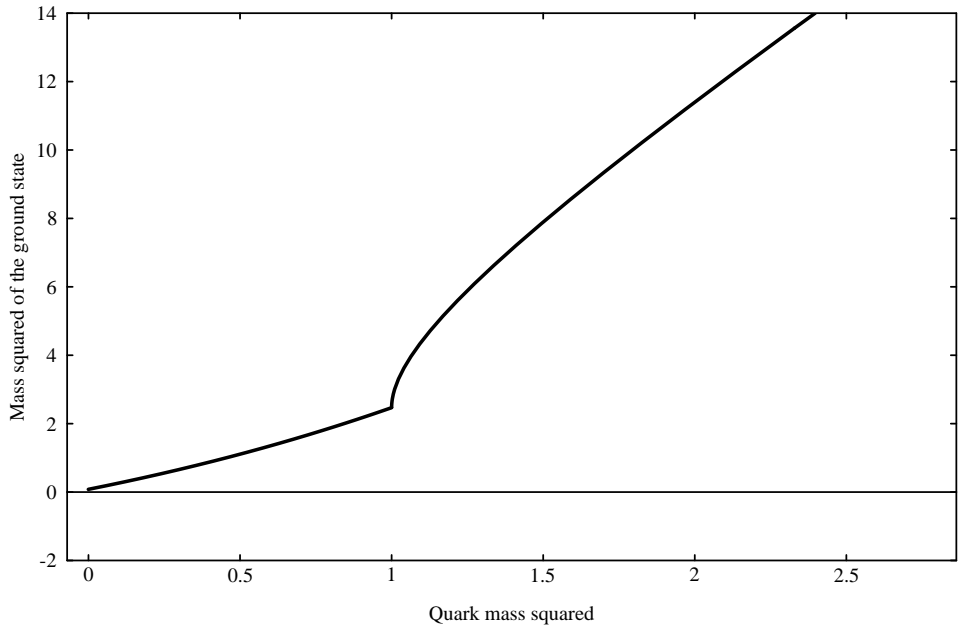

**FIGURE 1** Mass squared of the ground state meson vs. quark mass squared. Both axes are in units of $g^2 N/\pi$. The weak coupling regime is to the right of the point $m^2 = 1$, the strong coupling regime to the left of it.

In conclusion, the present work develops large-$N$ scalar quantum chromodynamics for light and strongly bound quarks. The wave functions introduced here and the techniques developed in order to compute with them make it now possible to do calculations in the entire range of quark masses. They can also be used in the full theory involving both scalar *and* fermion quarks in the strong coupling regime. As I will report in a forthcoming publication, I obtain massless composite fermions in that theory. **Footnote**

$^\star$ This leaves out the functions which, while going to zero at the boundaries, do so slower than any power. An example is provided by $\bigl(\ln x(1-x)\bigr)^{-1}$. This possibility is the reason for writing $x^0$ and $(1-x)^0$ elsewhere in this article. While I cannot exclude this possibility, in the present work I shall discard it.

## References

[1] G. 't Hooft, *Nucl. Phys.* **B75** (1974) 461.
[2] W.A.Bardeen, R.B.Pearson, and E. Rabinovici, *Phys. Rev.* **D21** (1980) 1037.
[3] S.S. Shei and H.S. Tsao, *Nucl. Phys.* **B141** (1978) 445.
[4] T. Tomaras, *Nucl. Phys.* **B163** (1980) 79.
[5] H. Wilf and D. Zeilberger, *Invent. Math.* **108** (1992) 575.